\documentclass[12pt]{article}
\pdfoutput=1
\usepackage{subfigure}
\usepackage{amssymb,amsmath}
\usepackage{graphicx}
\usepackage{color}
\usepackage[colorlinks=true
,urlcolor=blue
,citecolor=blue
,linkcolor=blue
,pagecolor=blue
,linktocpage=true
,pdfproducer=medialab
]{hyperref}
\usepackage[a4paper,width=16.8cm]{geometry}
\makeatletter \renewcommand{\@dotsep}{10000} \makeatother
\usepackage{appendix}

\begin{document}

\begin{center}

 {\Large\bf 
 Addressing Infinities in the L{\'e}vy-Leblond Hamiltonian
 } \vspace{1cm}

{   Muhammad Adeel Ajaib\footnote{ E-mail: adeel@udel.edu}}

{\baselineskip 20pt \it
Department of Physics, Coastal Carolina University, Conway, SC, 29526  } \vspace{.5cm}

{\baselineskip 20pt \it
   } \vspace{.5cm}

\setcounter{footnote}{0}
\vspace{1.5cm}
\end{center}

\begin{abstract}

We attempt to shed light on the following question: What happens to the negative energy states when we take the non-relativistic limit of the Dirac equation? The L{\'e}vy-Leblond equation is the non-relativistic limit of the Dirac equation and describes fermions in the non-relativistic limit. The L{\'e}vy-Leblond equation includes singular matrices and an attempt to write the Hamiltonian appears to show that the negative energy states are ``buried'' under an infinity. We attempt to isolate the infinite energy states and also present an equivalent way of viewing the Schrodinger dispersion relation.  We propose that the L{\'e}vy-Leblond equation can also be seen as resulting from the contribution of enhanced Lorentz violating terms to the Dirac equation.

\end{abstract}

\newpage

\section{Introduction}\label{intro}

Infinities frequently appear in quantum field theories and approaches to tackle these are typically referred to as renormalization methods. The aim of this article is to address the infinities that  arise in the L{\'e}vy-Leblond Hamiltonian. The L{\'e}vy-Leblond equation \cite{LevyLeblond:1967zz, Ajaib:2015uha} results from the non-relativistic limit of the Dirac equation and describes fermions in the non-relativistic limit. The L{\'e}vy-Leblond equation involves singular matrices and an attempt to write the Hamiltonian of this equation leads to two states that have finite 
energies ($E=p^2/2m$) whereas the other two are infinite ($E=m/ \epsilon - p^2/2m$, $\epsilon \rightarrow 0$) \cite{Ajaib:2017}.

The Dirac equation has negative energy states which are interpreted as antiparticles. 
The motivation of this article is the following: What happens to these negative energy states in the transition from the Dirac equation to the L{\'e}vy-Leblond equation. The negative energy states appears to be ``buried'' under an infinity. In an attempt to isolate these infinities we suggest an equivalent way of interpreting the Schrodinger dispersion relation. This interpretation implies that the the L{\'e}vy-Leblond Lagrangian is obtained by including certain Lorentz violation terms to the Dirac Lagrangian.

\section{The L{\'e}vy-Leblond Equation and Infinities }

In this section we describe how infinities arise in the L{\'e}vy-Leblond Hamiltonian and in the following section we shall address how to isolate them. The Dirac equation is given by ( $\hbar=c=1$)
\begin{eqnarray}
(i   \gamma^\mu \partial_\mu  -  m) \psi =0
\label{DE}
\end{eqnarray}
or in momentum space 
\begin{eqnarray}
( \gamma_0 E- \gamma_i p_i   -  m) \psi =0
\end{eqnarray}
where $\gamma_\mu$ are the Dirac matrices. The non-relativistic limit of the Dirac equation yields the L{\'e}vy-Leblond equation \cite{Ajaib:2015eer}:
\begin{eqnarray}
\left( \eta_1 E' - \gamma_i p_i + \eta_2 m \right) u =0
\label{eq:nrde}
\end{eqnarray}
Here $\eta_{1}=(\gamma_0+ I)/2$ and $\eta_{2}=(\gamma_0 - I)$, and the above equation yields the dispersion relation of a non-relativistic particle ($E'=p_i^2/2m$). The matrices $\eta_{1,2}$ are singular, hermitian and $\eta_1\eta_2=0$. Note that in \cite{Ajaib:2015eer} the non-relativistic limit of the Dirac equation was analyzed with the $i\gamma_5$ mass term. The matrices obtained in \cite{Ajaib:2015eer} were non-hermitian and singular. The Hamiltonian corresponding to the L{\'e}vy-Leblond equation is given by
\begin{eqnarray}
H_L=\eta^{\prime -1} (-i \gamma_i \partial_i-m \eta^{\dagger})
\label{eq:hamiltonian}
\end{eqnarray}
where we have chosen $\eta^\prime=\eta-\epsilon \eta^\dagger$. In the limit $\epsilon \rightarrow 0$, two of the eigenvalues of the Hamiltonian in (\ref{eq:hamiltonian}) are finite where as two approach infinity
\begin{eqnarray}
E_{1,2} &=& \frac{\vec{p}^{\ 2}}{2m} \\
E_{3,4} &=& -\frac{\vec{p}^{\ 2}}{2m}+\frac{m}{\epsilon}
\end{eqnarray}
We can see that the negative energy states appear to be ``hidden'' under an infinity. For the negative energy states we can also define the renormalized energy as 
\begin{eqnarray*}
E^\prime_{3,4}=E_{3,4}-\frac{m}{\epsilon}= -\frac{\vec{p}^{\ 2}}{2m} 
\end{eqnarray*}

\section{Isolating the Negative Energy States}

In this section we present the Hamiltonian that results from attempting to isolate the infinities encountered in the L{\'e}vy-Leblond equation. We first discuss how this exclusion appears for scalars by suggesting a rescaling of the Dirac dispersion relation as follows:
\begin{eqnarray}
E &=& \pm \sqrt{p^2+m^2} \\
\pm E&=&  \frac{p^2}{2m}+m -\frac{p^4}{8 m^3}+\frac{p^6}{16 m^5}+ \ldots \\
\pm E-E^\prime  &=&  \frac{p^2}{2m}
\end{eqnarray}
where
\begin{eqnarray}
E^\prime &=& m-\frac{p^4}{8 m^3}+\frac{p^6}{16 m^5}+ \ldots \\
&=&  \pm \sqrt{p^2+m^2}-  \frac{p^2}{2m}
\end{eqnarray}
Note that $-\infty \le E^\prime \le m$. Next, we will discuss how the above approach appears for fermions by considering  the Dirac Hamiltonian ($H_D=\gamma_0 \gamma_i p_i +m \gamma_0$) and the L{\'e}vy-Leblond Hamiltonian (\ref{eq:hamiltonian}). Subtracting the two Hamiltonians in the above manner can isolate the infinite energy states:
\begin{eqnarray}
H^\prime = H_D-H_L
\end{eqnarray}
Here the eigenvalues of the Hamiltonian $H^\prime$ are ($m$, $m$, $-m-m/\epsilon$,$-m-m/\epsilon$). We can interpret the above equation as follows. The contribution of these states with positive rest mass energy ($E=m$) and infinite negative energy ($E=-m-m/\epsilon$) to the L{\'e}vy-Leblond Hamiltonian leads to the Dirac Hamiltonian ($ H_D=H_L+H^\prime$). 
Further insights into this approach can be obtained by considering the Lagrangian approach.
The Dirac Lagrangian (${\cal L}_{D}$) and the  Lagrangian corresponding to the L{\'e}vy-Leblond equation  (${\cal L}_{L}$) are given by
\begin{eqnarray}
{\cal L}_{D}=i \bar{\psi} \gamma^\mu \partial_\mu \psi - m  \bar{\psi}\psi
\end{eqnarray}
\begin{eqnarray}
{\cal L}_{L}=i \bar{\psi} \eta^\mu \partial_\mu \psi + m \bar{\psi} \eta_2 \psi
\end{eqnarray}
where  $\eta^\mu=(\eta^0,\eta^i)=(\eta_1,\gamma^i)$. The Lagrangian corresponding to the infinite energy states can be written as 
\begin{eqnarray}
{\cal L}^\prime={\cal L}_{D}-{\cal L}_{L}
\end{eqnarray}
Here ${\cal L}^\prime$ includes terms that violate Lorentz invariance. These terms, when added to the Dirac Lagrangian, yield the Lagrangian corresponding to (\ref{eq:nrde}), i.e.,
\begin{eqnarray}
{\cal L}^\prime=i \bar{\psi} \left( \frac{\gamma_0-I}{2} \right) \partial_0 \psi - m \bar{\psi} \gamma_0 \psi
\end{eqnarray}
Therefore in light of the above approach we can say that the L{\'e}vy-Leblond Lagrangian results from the contribution of certain Lorentz violating terms in the Dirac Lagrangian.

\section{Conclusion} \label{conclude}

We attempted to better understand the infinities that arise in the L{\'e}vy-Leblond Hamiltonian. We suggest two possible ways of viewing the L{\'e}vy-Leblond equation. One is that it is the non-relativistic limit of the Dirac equation. Another equivalent way is that it is obtained from the contribution of enhanced Lorentz violating terms to the Dirac equation.


\end{document}